\def\draftversion{false}
\RequirePackage{ifthen}
\ifthenelse{\equal{\draftversion}{true}}{
  \documentclass[aps,showpacs,prl,galley,superscriptaddress,
                 longbibliography]{revtex4-1}
}{
  \documentclass[aps,showpacs,prl,reprint,superscriptaddress,
                 longbibliography]{revtex4-1}
}

\usepackage[colorlinks=true,linkcolor=blue,citecolor=blue,urlcolor=blue]{hyperref}

\usepackage{hyperref}
\usepackage{setspace} %to set 1.5 spacing in text
\usepackage{graphicx}
\usepackage{amsmath}
\usepackage[usenames]{color}
\usepackage{amsmath}
\usepackage{amssymb}
\usepackage{verbatim}
\usepackage{latexsym}
\usepackage{enumerate} % to change the numbering of 'enumerate'
\usepackage{bm} % for bold face mathematical symbols
\usepackage[caption=false]{subfig}
\usepackage{soul}  % \st    for strike-out

\ifthenelse{\equal{\draftversion}{true}}{
  \marginparwidth 2.7in
  \marginparsep 0.5in
  \newcounter{comm} % counter for commentaries
  \def\commnext{\stepcounter{comm}}
  \def\commtext{{\bf\color{blue}[\arabic{comm}]}}
  \def\commmar{{\bf\color{blue}[\arabic{comm}]}}
  \def\dvm#1{\commnext\marginpar{\small DV\commmar: #1}\commtext}
  \def\tmm#1{\commnext\marginpar{\small BM\commmar: #1}\commtext}
  \def\mlab#1{\marginpar{\small\bf #1}}
  
}{
  \def\dvm#1{}
  \def\tmm#1{}
  \def\mlab#1{}
  
}
\begin{document}

\title{Antiferroelectric topological insulators in orthorhombic $A$MgBi compounds ($A=$ Li, Na, K)}

\author{Bartomeu Monserrat}
\email{monserrat@physics.rutgers.edu}
\affiliation{Department of Physics and Astronomy, Rutgers University,
  Piscataway, New Jersey 08854-8019, USA}
\affiliation{TCM Group, Cavendish Laboratory, University of Cambridge,
  J.\ J.\ Thomson Avenue, Cambridge CB3 0HE, United Kingdom}
\author{Joseph W. Bennett}
\affiliation{Department of Physics and Astronomy, Rutgers University,
  Piscataway, New Jersey 08854-8019, USA}
\author{Karin M. Rabe}
\affiliation{Department of Physics and Astronomy, Rutgers University,
  Piscataway, New Jersey 08854-8019, USA}
\author{David Vanderbilt}
\affiliation{Department of Physics and Astronomy, Rutgers University,
  Piscataway, New Jersey 08854-8019, USA}

\date{\today}

\begin{abstract}
We introduce antiferroelectric topological insulators as a new class of functional materials in which an electric field can be used to control topological order and induce topological phase transitions. Using first principles methods, we predict that several alkali-MgBi orthorhombic members of an $ABC$ family of compounds are antiferroelectric topological insulators. We also show that epitaxial strain and hydrostatic pressure can be used to tune the topological order and the band gap of these $ABC$ compounds. Antiferroelectric topological insulators could enable precise control of topology using electric fields, enhancing the applicability of topological materials in electronics and spintronics.
\end{abstract}


\maketitle

% Electric field control of topological insulators
Topological insulators and related materials~\cite{z2_topological_insulators,topological_insulators_review,topological_insulators_review2} exhibit unusual properties such as robust edge currents and spin-momentum locking. These unconventional properties have led to a plethora of proposals for the use of topological materials in fundamental research spanning from magnetic monopoles~\cite{ti_magnetic_monopole} to Majorana fermions~\cite{ti_majorana_fermions}, and in applications such as fault-tolerant quantum computers. In this context, an innovative route towards the control of topological order is by means of electric fields~\cite{efield_sb2te3_films,efield_sb2te3_exp,murakami_ferro_topo_superlattice,efield_ti_zunger,narayan_abc_feti,rappe_feti,picozzi_topo_hyperferroelectrics}. 

\begin{figure}[b]
\centering
\subfloat[][]{\includegraphics[scale=0.30]{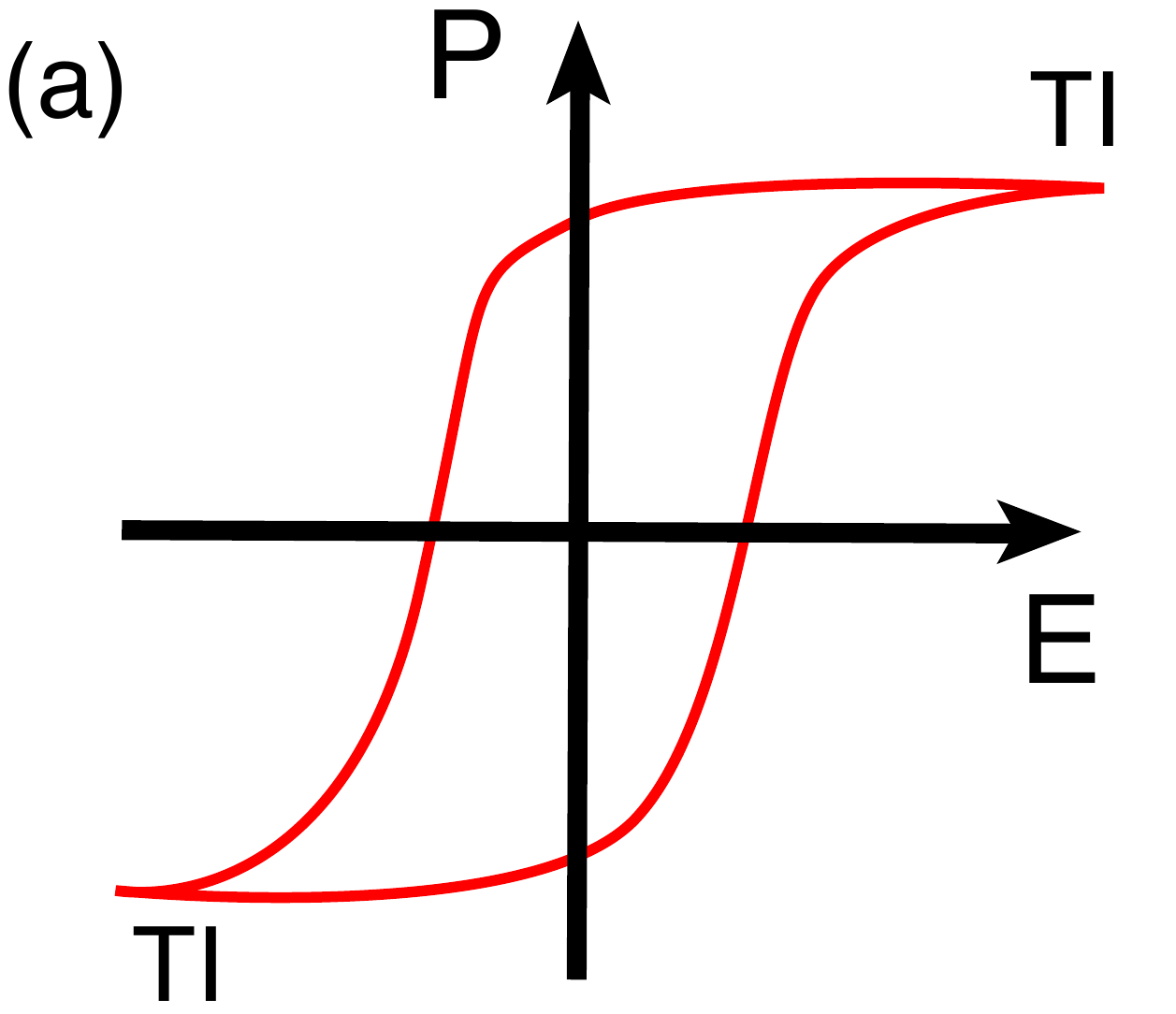}\label{subfig:feti}}
\hspace{0.3cm}
\subfloat[][]{\includegraphics[scale=0.30]{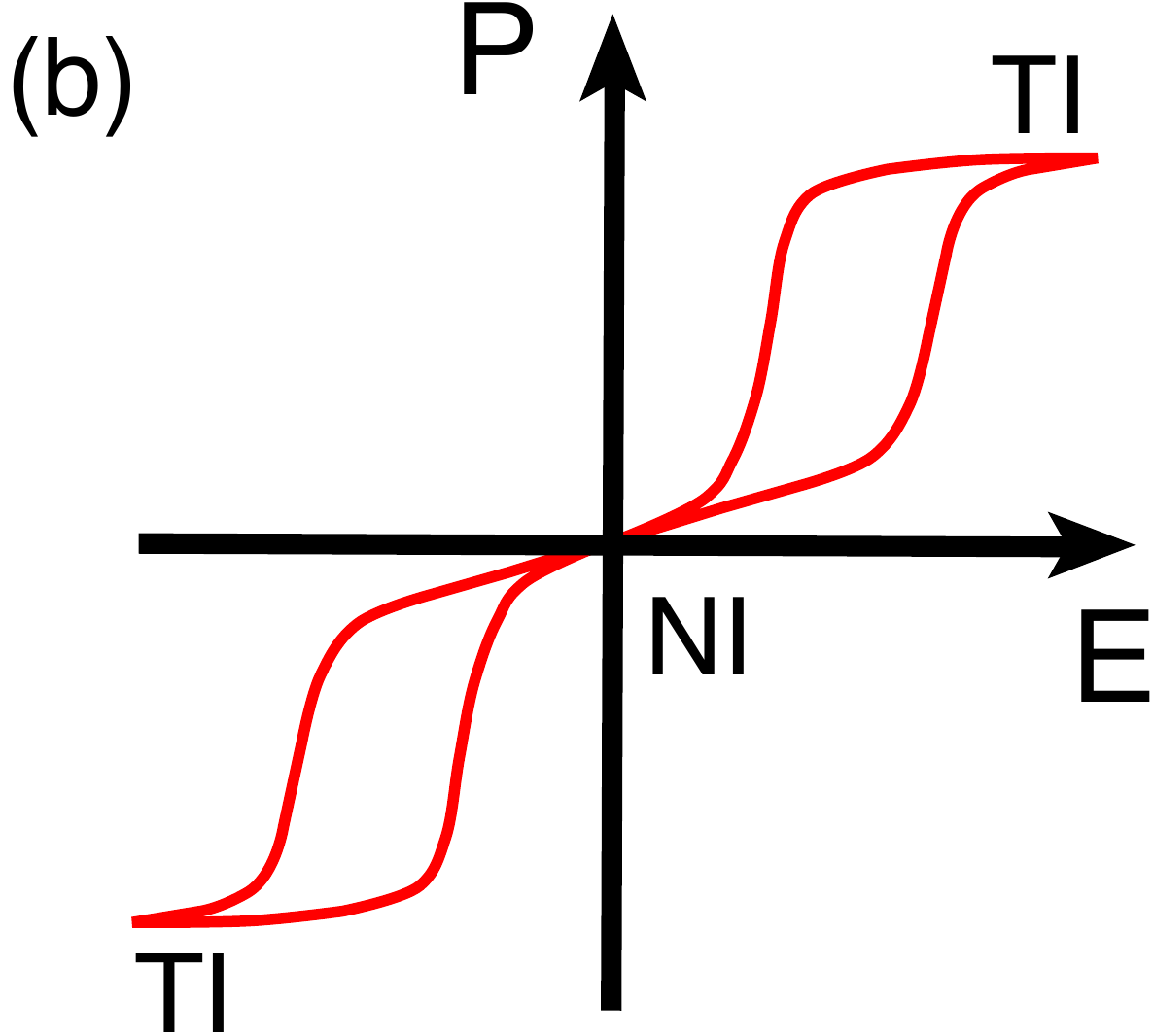}\label{subfig:afti}}
\caption{Schematic representation of (a) a ferroelectric topological insulator and (b) an antiferroelectric topological insulator with an antipolar normal insulator (NI) state and two polar topological insulator (TI) states, referred to as type-I antiferroelectric topological insulator in the text.}
\label{fig:afti}
\end{figure}

% Overview of previous work on electric field control of topological insulators
A promising direction to control topological order with an electric field is to utilize ferroelectric materials that harbor topological states. Ferroelectrics exhibit two states of opposite polarity ($\mathbf{P}$) that can be controlled by an electric field ($\bm{\mathcal{E}}$) as shown schematically in Fig.~\ref{subfig:feti}.  %as shown schematically in Fig.~\ref{subfig:feti}. 
The two polarization states in these ferroelectric topological insulators could be used, for example, to create spin-selected collimated electron beams~\cite{rappe_feti} or to control the spin texture around the Dirac cones of the surface states~\cite{picozzi_topo_hyperferroelectrics}. Several materials have been proposed to exhibit these properties: superlattice combinations of the ferroelectric material GeTe and the topological insulator Sb$_2$Te$_3$ exhibit electric field control of topological order~\cite{murakami_ferro_topo_superlattice}; strained LiZnSb and CsPbI$_3$ are candidate ferroelectric topological insulators~\cite{narayan_abc_feti,rappe_feti}; and ferroelectricity and topological order coexist in some $ABC$ hyperferroelectrics, which could be candidates for thin-film applications~\cite{picozzi_topo_hyperferroelectrics}.

% Statement of what we do: antiferroelectric topological insulator
In this work we explore the possibility of controlling topological order in antiferroelectric materials, which exhibit an antipolar ground state and two polar states that can be accessed using an electric field, as shown schematically in Fig.~\ref{subfig:afti}. In ferroelectric materials, the two polar states are related by inversion symmetry, imposing the same topological character on both. Antiferroelectrics are more flexible, as inversion symmetry also dictates that the two polar states have the same topological character, but the antipolar state could belong to a different topological class. 
We thus define an antiferroelectric topological insulator (AFTI) as an antiferroelectric material in which at least one of the states is a topological insulator. There can be three different types of AFTIs, depending on which states are topological insulators:
(i) a type-I AFTI with an antipolar normal state and polar topological states, (ii) a type-II AFTI with an antipolar topological state and polar normal states, and (iii) a type-III AFTI with antipolar and polar topological states.
Electric fields could be used to drive topological phase transitions in AFTIs of types I and II. Of these, AFTIs of type I are perhaps the most interesting as their topological polar states would exhibit the properties of ferroelectric topological insulators. The schematic of Fig.~\ref{subfig:afti} shows an example of an AFTI of type I.

% Description of system class and justification
We study potential AFTIs in the $ABC$ family of materials which includes compounds that are proposed to be ferroelectric~\cite{abc_ferroelectrics}, antiferroelectric~\cite{abc_antiferroelectrics}, and hyperferroelectric~\cite{abc_hyperferroelectrics}. 
The antiferroelectric $ABC$ compounds have a nonpolar reference structure of hexagonal $P6_3/mmc$ space group with energy-lowering distortions to an antipolar orthorhombic $Pnma$ structure and to a polar hexagonal $P6_3mc$ structure. The $Pnma$ antipolar structure is the ground state, and there is a small energy barrier to the $P6_3mc$ polar structure~\cite{abc_antiferroelectrics}.
 We adopt a convention in which the $B$ and $C$ atoms form the hexagonal layers of the $P6_3/mmc$ structure, and the $A$ atoms are arranged in the stuffing sites in between. 
%Bennett and co-workers have proposed a range of compounds in which the $Pnma$ antipolar structure has the lowest energy, with a small energy barrier to the $P6_3mc$ polar structure, which suggests the possibility of using these materials as antiferroelectrics~\cite{abc_antiferroelectrics}. 
The large number of possible $ABC$ compounds with these characteristics make them an excellent platform to search for the coexistence of antiferroelectricity and other types of order. 

% Search strategy and target compounds
Topological insulators exhibit band inversion driven by the spin-orbit interaction. Since the strength of the spin-orbit interaction increases with the atomic number, we focus the search for topological materials in $ABC$ compounds to those containing heavy elements. Furthermore, the energy scale associated with spin-orbit coupling (of up to about $1$~eV) needs to be comparable to the band gap in order to induce a band inversion, and therefore we focus on materials with band gaps below $1$~eV. We use these criteria to search over the $ABC$ antiferroelectrics reported in Ref.~\cite{abc_antiferroelectrics}, and we indentify three compounds, LiMgBi, NaMgBi, and KMgBi, that are promising candidates as AFTIs. We refer collectively to these three compounds as $A$MgBi (for $A=$Li, Na, K). 

\paragraph{First-principles calculations.}
% DFT structural calculations
Our calculations are based on density functional theory (DFT) as implemented in the {\sc vasp} package~\cite{vasp1,vasp2,vasp3,vasp4}. We use the local density approximation (LDA) to the exchange correlation functional~\cite{PhysRevLett.45.566,PhysRevB.23.5048} and the projector augmented wave method~\cite{paw_original,paw_us_relation} with an energy cut-off of $300$~eV and a $\mathbf{k}$-point grid size of $8\times8\times8$ to sample the Brillouin zone (BZ). All calculations include spin-orbit coupling. We also perform selected calculations using the HSE hybrid functional~\cite{hse06_functional,hse06_functional_erratum} to estimate the robustness of the first-principles predictions.

The lattice vectors and internal coordinates are optimized until the forces on all atoms are smaller than $0.001$~eV/\AA\@, and the pressure is within $0.01$~GPa of the target value when applying hydrostatic pressure. In-plane epitaxial strain on a hexagonal lattice is imposed by constraining the in-plane lattice constants and allowing the out-of-plane lattice constant and the internal coordinates to fully relax. Strains are reported using the hexagonal $P6_3mc$ equilibrium structure as reference, with in-plane lattice constants equal to $4.68$~\AA\@, $4.86$~\AA\@, and $5.05$~\AA\@ for LiMgBi, NaMgBi, and KMgBi, respectively. The $Pnma$ equilibrium structure differs from any epitaxially strained $Pnma$ structure because the imposition of a hexagonal substrate forces the two in-plane lattice vectors of the orthorhombic structure to have the exact ratio $\sqrt{3}$, a condition that is not exactly obeyed in the unstrained structure.

% DFT band gap and Z2 calculations
We make a precise determination of the band gaps by Wannier interpolation of the band structure~\cite{wannier_rmp} using the {\sc wannier90} package~\cite{wannier90_2008,wannier90_2014}. We use the iterative approach of Ref.~\cite{liu_noncentrosymm_weyl}, initially sampling the electronic BZ with a coarse $\mathbf{k}$ grid (typically of size $40\times40\times40$), then determining the $\mathbf{k}_0$-point having the smallest gap, and then searching over a denser $\mathbf{k}$-point grid centered around the point $\mathbf{k}_0$. This procedure is iterated until convergence is reached. 

Once we have determined that a phase is insulating, we calculate the topological $Z_2$ invariant, which in $3$ dimensions is determined by four indices $(\nu_0;\nu_1\nu_2\nu_3)$, where the first index $\nu_0$ is the strong topological index and determines whether the material is a strong topological insulator ($\nu_0=1$) or not ($\nu_0=0$), and the other indices are the weak topological indices that determine the reciprocal space planes in which band inversions occur~\cite{3d_ti_fu,3d_ti_moore,3d_ti_roy}. We calculate the topological indices by following the adiabatic pumping of Wannier charge centers over the BZ~\cite{z2_index_calculation} as implemented in the {\sc z2pack} code~\cite{z2pack}; an approach that is appropriate for systems without inversion symmetry. 

% Ambient conditions
\begin{table}
  \setlength{\tabcolsep}{2pt} % Default is 6pt
  \caption{Energy difference $\Delta E$ between the $P6_3mc$ structure and the $Pnma$ structure in meV per formula unit (f.u.), and the band gaps $E_{\mathrm{g}}^{P6_3mc}$ and $E_{\mathrm{g}}^{Pnma}$ of the $P6_3mc$ and the $Pnma$ structures in meV. All results correspond to the equilibrium structures and the use of the LDA functional.}
  \label{tab:ambient}
  \begin{ruledtabular}
  \begin{tabular}{c|ccc}
            &  $\Delta E$~(meV/f.u.)   &   $E_{\mathrm{g}}^{P6_3mc}$~(meV)   & $E_{\mathrm{g}}^{Pnma}$~(meV)   \\
  \hline
  LiMgBi &  $59$  & $66$  & $12$  \\
  NaMgBi &  $122$ & $85$  & $22$  \\
  KMgBi  &  $239$ & $96$  & $16$  \\
\end{tabular}
\end{ruledtabular}
\end{table}

\paragraph{Equilibrium structures.}
The $Pnma$ structure is lower in energy compared to the $P6_3mc$ structure in all three $A$MgBi materials, as shown in Table~\ref{tab:ambient}. Using the LDA functional, we predict that all three materials in both the $P6_3mc$ and $Pnma$ structures are strong topological insulators with topological indices $(\nu_0;\nu_1\nu_2\nu_3)=(1;000)$. The minimum band gaps, shown in Table~\ref{tab:ambient}, occur near the $\Gamma$ point, with states dominated by bismuth and magnesium. These results indicate that the $A$MgBi compounds are candidate topological insulators. Furthermore, as they are also candidate antiferroelectrics, these materials become candidate AFTIs of type III in which the the antipolar and both polar states have the same $Z_2$ indices.

 %Finally, we note that a polymorph of LiMgBi, namely MgLiBi in which the Li and Mg atoms are exchanged, is more stable than LiMgBi in our calculations.

\begin{figure}[b]
\centering
\includegraphics[width=0.45\textwidth]{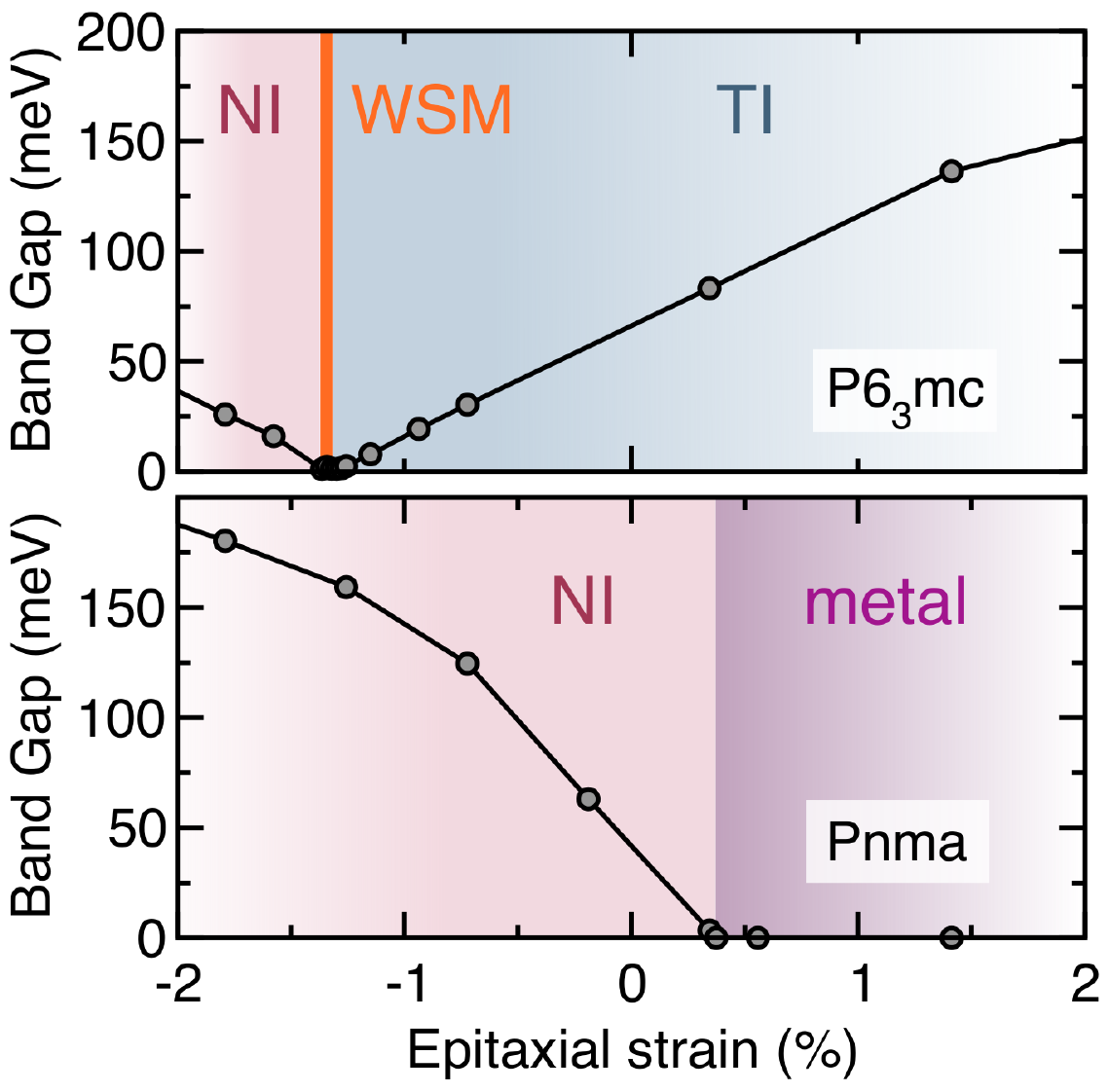}
\caption{Band gap of the LiMgBi compound of polar $P6_3mc$ (top) and antipolar $Pnma$ (bottom) structures as a function of epitaxial strain, calculated using the LDA functional. %together with their topological character. 
The $P6_3mc$ struture exhibits regions in which it is a topological insulator (TI), a Weyl semimetal (WSM), and a normal insulator (NI), whereas the $Pnma$ structure exhibits metallic and normal insulator phases. Note that the $Pnma$ structure at zero strain is distinct from the equilibrium $Pnma$ structure due to the $\sqrt{3}$ ratio that the in-plane lattice parameters must obey.}
\label{fig:gap_lmb}
\end{figure}

\begin{figure}
\centering
\includegraphics[width=0.45\textwidth]{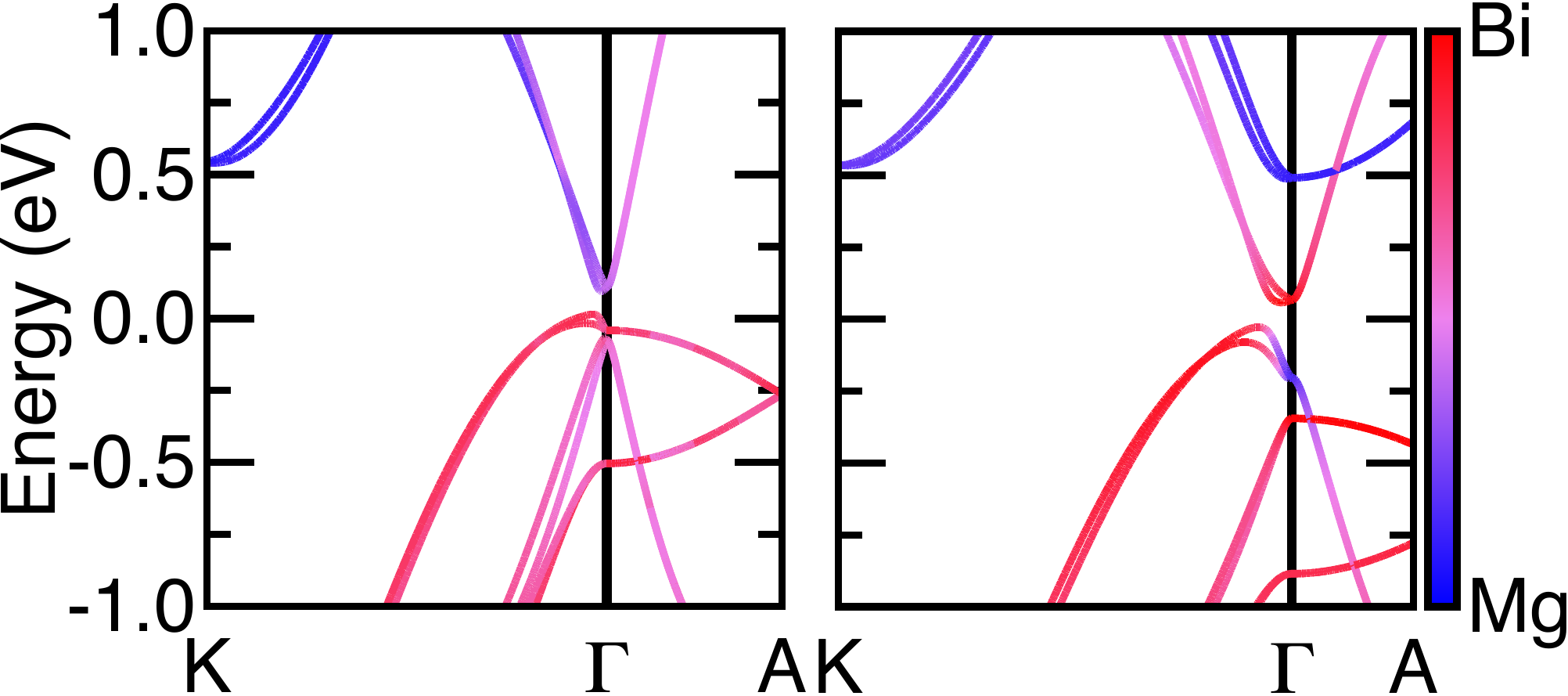}
\caption{Band structure of LiMgBi in the polar $P6_3mc$ structure at strains of $-2.9\%$ (left) corresponding to a normal insulator, and $+0.3\%$ (right) corresponding to a topological insulator.}
\label{fig:bands}
\end{figure}

\paragraph{Epitaxial strain.} 
% Epitaxial strain -- LiMgBi
The equilibrium structures are type III candidate AFTIs, but it would be desirable to find AFTIs of types I and II as well. With this aim, we further investigate how the relative energies between the $Pnma$ and $P6_3mc$ structures, as well as their topological character, can be tuned by means of epitaxial strain. Using the LDA functional, we find that the unstrained $P6_3mc$ structure of all three compounds corresponds to a topological insulator, and compressive epitaxial strain drives them through a topological phase transition to a normal insulator via an intermediate Weyl semimetal phase, as shown in Fig.~\ref{fig:gap_lmb} for LiMgBi. The intermediate semimetallic phase is a universal feature of a topological phase transition in materials without inversion symmetry~\cite{murakami_weyl_semimetal,liu_noncentrosymm_weyl}. In Fig.~\ref{fig:bands} we show the band structure of polar $P6_3mc$ LiMgBi around the $\Gamma$ point for epitaxial strains of $-2.86\%$, corresponding to a normal insulator, and $+0.34\%$, corresponding to a topological insulator, with gaps of $93$ and $94$~meV respectively. The bands exhibit Rashba splitting, the strength of which can also be tuned by means of epitaxial strain. The projection on the bismuth orbitals demonstrates the band inversion associated with the topological phase transition. 

In LiMgBi, the epitaxially strained $Pnma$ structure is a normal insulator under compressive strain, and becomes a metal at about $0.34\%$ tensile strain (Fig.~\ref{fig:gap_lmb}). The fact that the epitaxial LiMgBi $Pnma$ structure is either a normal insulator or a metal, whereas the corresponding unstrained structure is a topological insulator, derives from the fixed $\sqrt{3}$ ratio that the two in-plane orthorhombic axes obey under epitaxial strain. The violation of this condition is largest in LiMgBi, reaching $8$\%, whereas it is $6$\% in NaMgBi, and almost perfectly obeyed in KMgBi. We have performed band gap and $Z_2$ index calculations of several LiMgBi structures linearly interpolating between the epitaxially strained and unstrained $Pnma$ structures, confirming that there is a gap closure and a topological phase transition from the unstrained topological insulator to the epitaxial normal insulator. 

% LMB -- AFTI
The LDA band gap results presented in Fig.~\ref{fig:gap_lmb} indicate that epitaxial LiMgBi is a candidate AFTI of type I with the polar states corresponding to a topological insulator, and the antipolar state to a normal insulator, the situation depicted in Fig~\ref{subfig:afti}. Moreover, the energy difference between the $Pnma$ and the $P6_3mc$ structures decreases from $59$~meV for the equilibrium configuration to $18$~meV at zero strain, and the difference decreases further with increasing compressive strain, promoting the possibility that an electric field can be used to switch between the polar and antipolar states. Finally, we note that a polymorph of LiMgBi, namely MgLiBi in which the Li and Mg atoms are exchanged, is more stable for compressive strains above $0.20\%$.
%, so that the potential synthesis of LiMgBi as an AFTI should focus in the epitaxial strain range from $-0.20\%$ to $+0.34\%$.

% Epitaxial strain -- NMB and KMB
In NaMgBi and KMgBi, the $Pnma$ structure under epitaxial strain does not exhibit a normal insulator regime, and therefore
an antipolar normal state cannot coexist with a polar topological state
in these materials.
Instead, in NaMgBi a polar normal insulator coexists with an antipolar topological insulator at strains between $-4.72\%$ and $-4.00\%$, which corresponds to a type-II AFTI. However, the band gap for both antipolar and polar states in that region is below $20$~meV, which may make experimental realization difficult. KMgBi under strain behaves analogously to NaMgBi, and the polar normal insulator coexists with the antipolar topological insulator at strains between $-4.95\%$ and $-3.96\%$, but again the band gaps are small. Furthermore, for both compounds the epitaxial strains required to reach the type II AFTI are possibly too large for experimental realization. Additional details of NaMgBi and KMgBi under strain are provided in the Supplemental Material.

\begin{figure}
\centering
\includegraphics[width=0.45\textwidth]{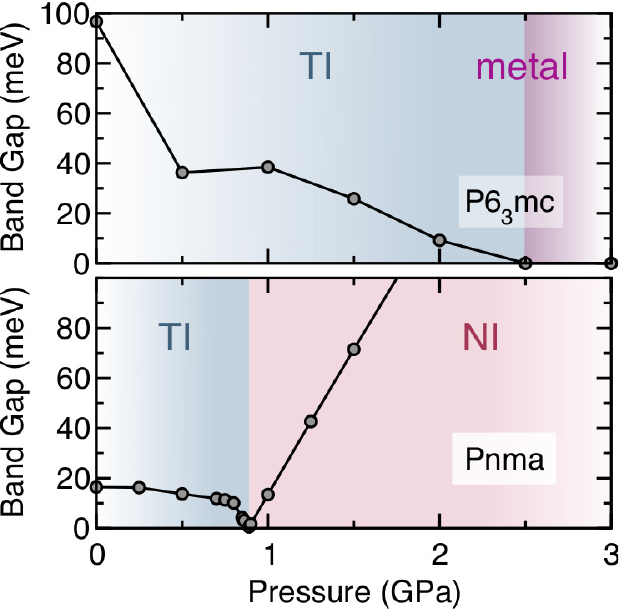}
\caption{Band gap of the KMgBi compound of polar $P6_3mc$ (top) and antipolar $Pnma$ (bottom) structures as a function of hydrostatic pressure calculated using the LDA functional. %together with their topological character. 
The $P6_3mc$ struture exhibits a topological insulator (TI) regime and a metallic regime, whereas the $Pnma$ structure exhibits topological insulator and normal insulator (NI) phases.}
\label{fig:gap_kmb}
\end{figure}

\begin{table*}[t]
  \setlength{\tabcolsep}{1pt} % Default is 6pt
  \caption{Candidate AFTIs from the $A$MgBi family of compounds based on LDA calculations (details for HSE calculations are provided in the text). Type I refers to normal antipolar and topological polar states, type II to topological antipolar and normal polar states, and type III to topological antipolar and polar states.}
  \label{tab:afti_summary}
  \begin{ruledtabular}
  \begin{tabular}{c|ccc}
            &  Equilibrium  &   Epitaxial strain   & Pressure   \\
  \hline
  LiMgBi &  --       & Type I ($-0.20$\% to $+0.34$\%)   & --  \\
  %LiMgBi &  --       & Type I                & --  \\
  %       &           & [$-0.20$\%,$+0.34$\%] &     \\
  NaMgBi &  Type III & Type II ($-4.72$\% to $-4.00$\%)  & --  \\
  %NaMgBi &  Type III & Type II                & --  \\
  %       &           & [$-4.72$\%,$-4.00$\%]  &   \\
  KMgBi  &  Type III & Type II ($-4.95$\% to $-3.96$\%)  & Type I ($0.89$~GPa to $2.50$~GPa)  \\
  %KMgBi  &  Type III & Type II                & Type I            \\
  %       &           & [$-4.95$,$-3.96$]\%  & [$0.89$,$2.50$]~GPa  \\
\end{tabular}
\end{ruledtabular}
\end{table*}

\paragraph{Hydrostatic pressure.}
% Hydrostatic pressure -- KMB
Under hydrostatic pressure and using the LDA functional, the $P6_3mc$ structure of NaMgBi and KMgBi undergoes a transition from a topological insulator into a metal, as shown in Fig.~\ref{fig:gap_kmb} for KMgBi. The $Pnma$ structure of NaMgBi and KMgBi undergoes a topological phase transition from a topological to a normal insulator, as shown for KMgBi in Fig.~\ref{fig:gap_kmb}. LiMgBi under hydrostatic pressure is higher in energy than the polymorph MgLiBi.

% KMB -- AFTI
The LDA band gap results of Fig.~\ref{fig:gap_kmb} suggest that KMgBi under hydrostatic pressure is also a candidate AFTI of type I (Fig.~\ref{subfig:afti}). In particular, the LDA functional prediction is that this phase should be observable at pressure between $0.89$~GPa and $2.50$~GPa. 
% Hydrostatic pressure - LMB and NMB
NaMgBi has a phase diagram similar to that of KMgBi, but there is no pressure range in which the normal antipolar and topological polar phases coexist, as shown in the Supplemental Material. 

\paragraph{Discussion.}
% Discussion
Overall, our first-principles calculations based on the LDA functional indicate that the $A$MgBi compounds are promising candidates as AFTIs, as summarized in Table~\ref{tab:afti_summary}.  Of these, the most promising candidates are LiMgBi under epitaxial strain in the range from $-0.20\%$ to $+0.34\%$, and KMgBi under hydrostatic pressure between $0.89$~GPa and $2.50$~GPa, which are AFTIs of type I with polar topological states and an antipolar normal state as depicted in Fig.~\ref{subfig:afti}. We have performed additional band gap calculations using the hybrid HSE functional for LiMgBi under epitaxial strain and KMgBi under pressure to assess our LDA-based predictions (see details in the Supplemental Material). We find that the HSE results lead to changes in the band gap that result in qualitatively similar phase diagrams but with significantly shifted phase boundaries. In particular, we find that the AFTI of type I in LiMgBi appears less favourable at the HSE level compared to the LDA level, as it is only present for epitaxial strains above $+2.3\%$. By contrast, we find that at the HSE level KMgBi is already an AFTI of type I under ambient conditions, without the need to apply any hydrostatic pressure. These results indicate that our predictions of the existence of AFTIs in the phase diagram are robust, but the locations of the phase boundaries in epitaxial strain or hydrostatic pressure depend on the functional used. Experimental exploration of these systems is desirable, with a focus on KMgBi under standard conditions or with moderate pressure, or LiMgBi under epitaxial strains in the range of $0-4\%$, as natural starting points. A signature of the topological phase could be the appearance of metallic surface states around the entire sample.

%The Fermi level would sit at the Dirac point on the surfaces with in-plane polarization, but above or below the Dirac point on the surfaces with out-of-plane polarization. The latter surface states provide a rich playground for spintronic applications due to spin-momentum locking of the associated electrons~\cite{rappe_feti}.

In LiMgBi and KMgBi AFTI phases of type I, an electric field could be used to induce a topological phase transition by switching between the antipolar and polar states. The electric field that makes the bulk electric enthalpy of the two phases equal is ${\cal E}=\Delta E/\Omega P$, where $P$ is the polarization of the polar phase and $\Omega$ is the cell volume. The energy differences $\Delta E$ between the antipolar and polar phases have an order of magnitude in the range $10$ to $100$~meV depending on the external conditions of strain or pressure, and the polarizations are about $0.4$~C/m~\cite{abc_antiferroelectrics}, leading to fields in the range $1.0\times10^7$ to $1.0\times10^8$~V/m. %We note that the lower fields are comparable to those used to switch standard antiferroelectrics~\cite{PbZrO3_antiferroelectric}.}

Although the field estimates above suggest that it should be feasible to use the $A$MgBi compounds as antiferroelectrics, we emphasize that in real materials switching is typically mediated by domain wall motion. This raises interesting questions for future research, as in an AFTI of type I or II, the domain walls mediating switching separate phases of distinct topological order.  As a result they are necessarily metallic, which may influence the switching dynamics in unusual and interesting ways. Free electrons screen static electric fields, and it might be necessary to use pulsed fields to switch an AFTI.

% Conclusions
\paragraph{Conclusions.}
We propose a new type of functional material, an antiferroelectric topological insulator, in which an electric field can be used to induce topological phase transitions and to control topological order. Using first-principles methods, we show that the orthorhombic $ABC$ compounds provide a rich playground in which to explore the existence of antiferroelectric topological insulators. In particular, the compounds $A$MgBi ($A=$Li, Na, K) exhibits an interesting interplay between antiferroelectric and topological properties as a function of both epitaxial strain and hydrostatic pressure, with LiMgBi under strain and KMgBi under pressure appearing as promising candidates for the experimental realization of an antiferroelectric topological insulator.

% Future work
%Further manipulation of the $A$MgBi compounds, such as the construction of superlattices, could lead to additional candidate antiferroelectric topological insulators. 

\acknowledgments

This work was funded by Office of Naval Research grants N00014-16-1-2591 and N00014-12-1-1040. B.M.~thanks Robinson
College, Cambridge, and the Cambridge Philosophical Society for a Henslow 
Research Fellowship.

% Mac
\bibliography{ti}

%%%%%%%%%% Merge with supplemental materials %%%%%%%%%%
\onecolumngrid
\clearpage
\begin{center}
\textbf{\large Supplemental Material for ``Antiferroelectric topological insulators in orthorhombic $A$MgBi compounds ($A=$ Li, Na, K)''}
\end{center}
%%%%%%%%%% Merge with supplemental materials %%%%%%%%%%
%%%%%%%%%% Prefix a "S" to all equations, figures, tables and reset the counter %%%%%%%%%%
\setcounter{equation}{0}
\setcounter{figure}{0}
\setcounter{table}{0}
\setcounter{page}{1}
\makeatletter
\renewcommand{\theequation}{S\arabic{equation}}
\renewcommand{\thefigure}{S\arabic{figure}}
\renewcommand{\bibnumfmt}[1]{[S#1]}
\renewcommand{\citenumfont}[1]{S#1}
%%%%%%%%%% Prefix a "S" to all equations, figures, tables and reset the counter %%%%%%%%%%
\section{N\lowercase{a}M\lowercase{g}B\lowercase{i} under epitaxial strain and hydrostatic pressure}

\begin{figure}[h]
\includegraphics[width=0.325\textwidth]{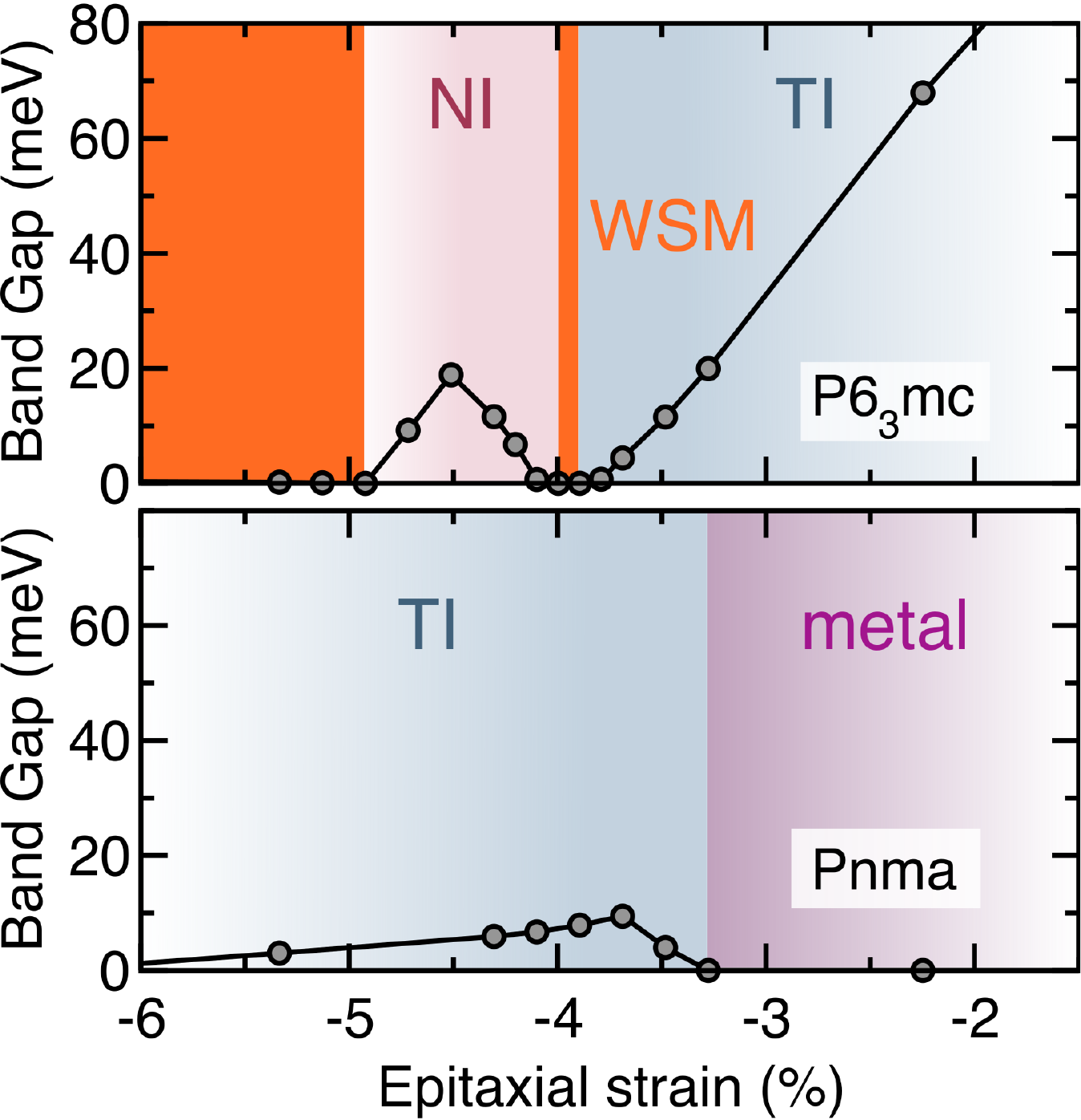}
\hspace{0.7cm}
\includegraphics[width=0.34\textwidth]{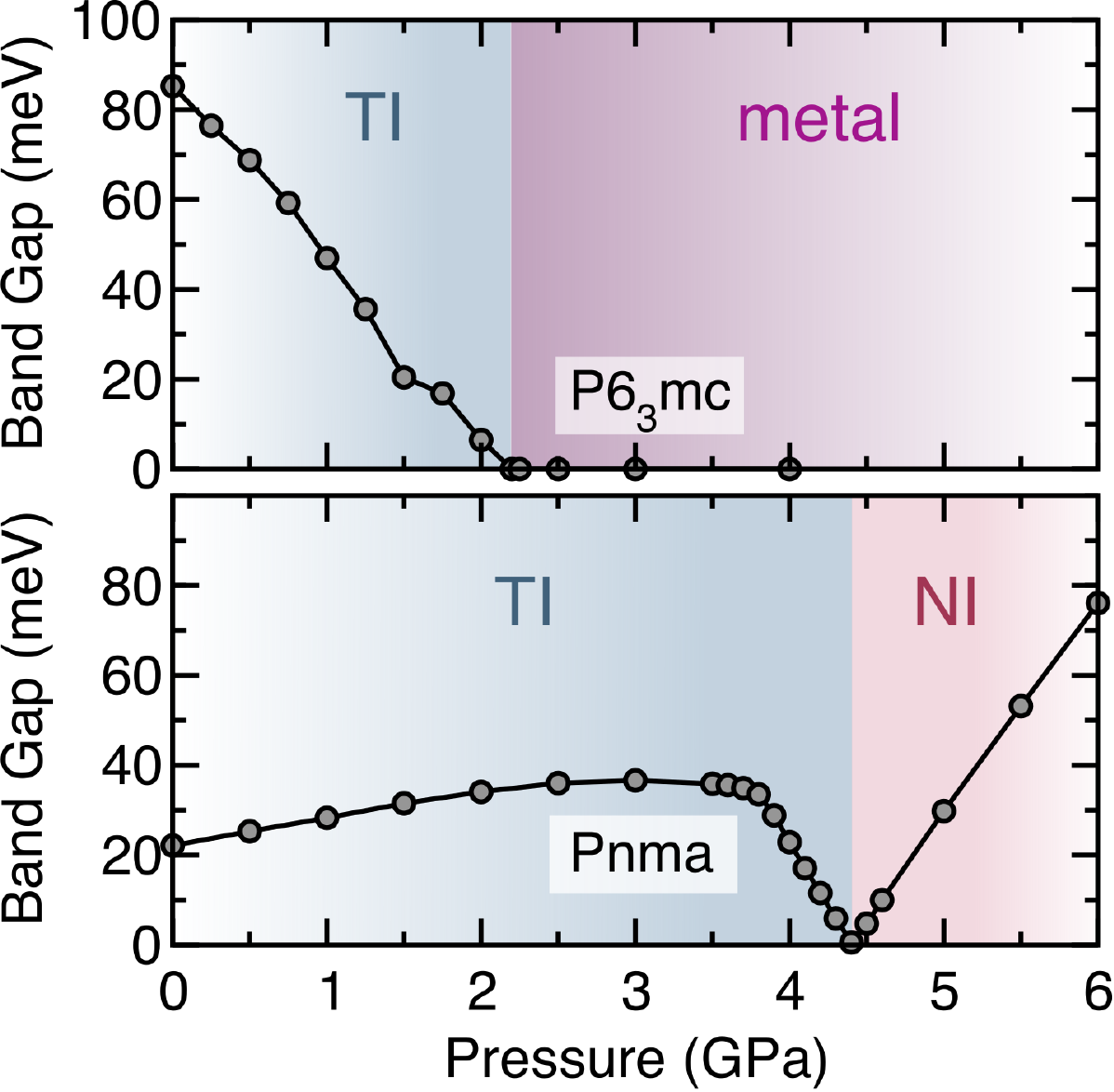}
\caption{Band gap of the NaMgBi compounds of polar $P6_3mc$ (top) and antipolar $Pnma$ (bottom) structures as a function of epitaxial strain (left) and hydrostatic pressure (right). The topological character of each phase is also indicated as normal insulator (NI), topological insulator (TI), Weyl semimetal (WSM), or metal.}
\label{fig:nmb}
\end{figure}

\nopagebreak

\section{KM\lowercase{g}B\lowercase{i} under epitaxial strain and hydrostatic pressure}

\begin{figure}[h]
\includegraphics[width=0.34\textwidth]{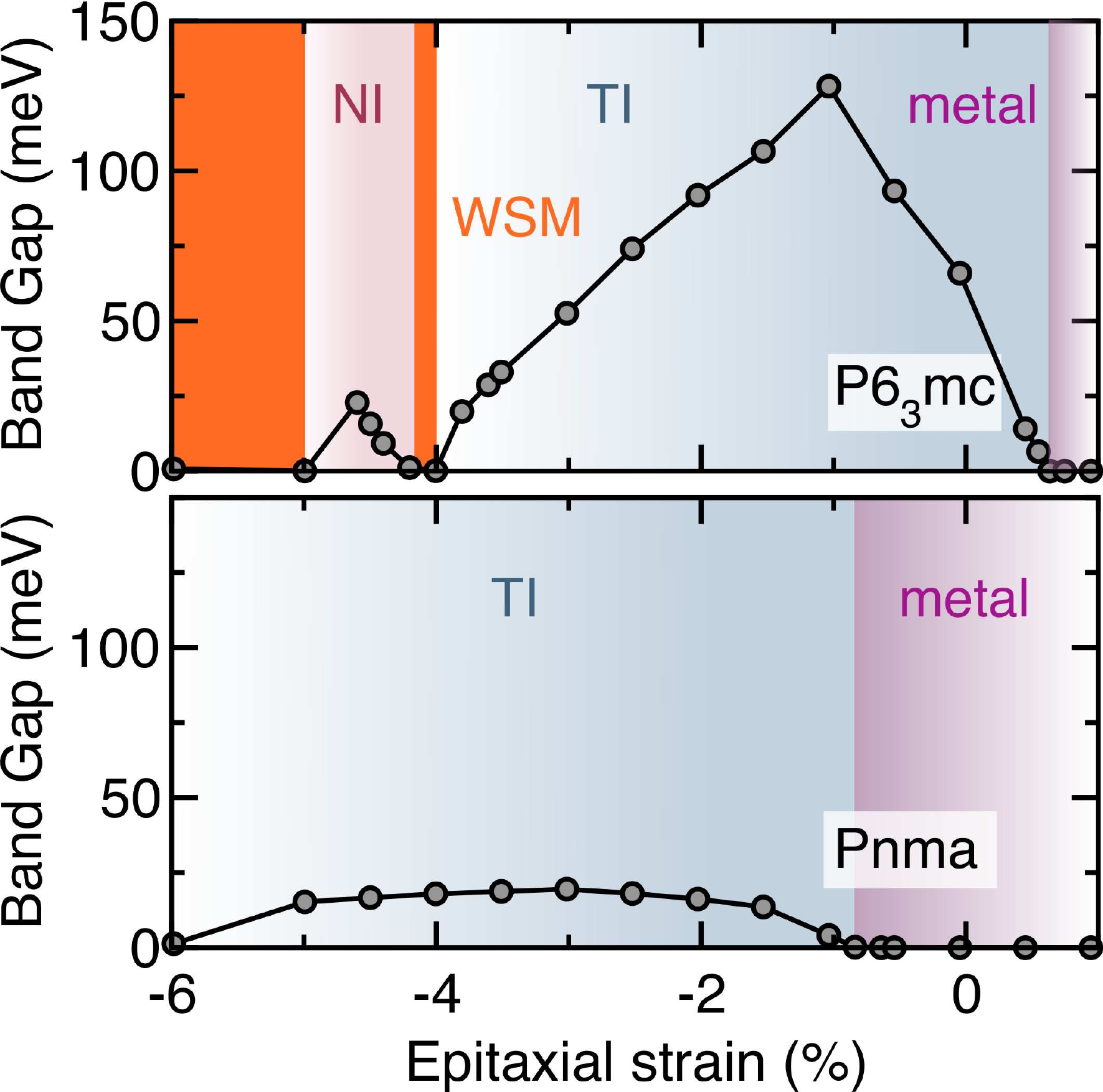}
\hspace{0.7cm}
\includegraphics[width=0.34\textwidth]{./gap_kmb_pressure.pdf}
\caption{Band gap of the KMgBi compounds of polar $P6_3mc$ (top) and antipolar $Pnma$ (bottom) structures as a function of epitaxial strain (left) and hydrostatic pressure (right). The topological character of each phase is also indicated as normal insulator (NI), topological insulator (TI), Weyl semimetal (WSM), or metal. The graph of hydrostatic pressure (right) is reproduced from the main manuscript for completeness.}
\label{fig:kmb}
\end{figure}

\section{Hybrid functional calculations}

The band gap results presented in the main text have been calculated using the local density approximation (LDA) to the exchange-correlation functional~\cite{PhysRevLett.45.566,PhysRevB.23.5048}. It is well-known that band gaps evaluated using semilocal functionals such as the LDA tend to underestimate the magnitude of the band gap for normal insulators. Topological properties are insensitive to the precise magnitude of the band gap, and can only change when the band gap closes. For this reason, we expect that the phase diagrams presented in the main text are qualitatively independent of the approximation to the exchange-correlation functional used, and only the precise location of the phase boundaries might change with different levels of theory.

\begin{figure}
\centering
\subfloat[][]{\includegraphics[scale=0.40]{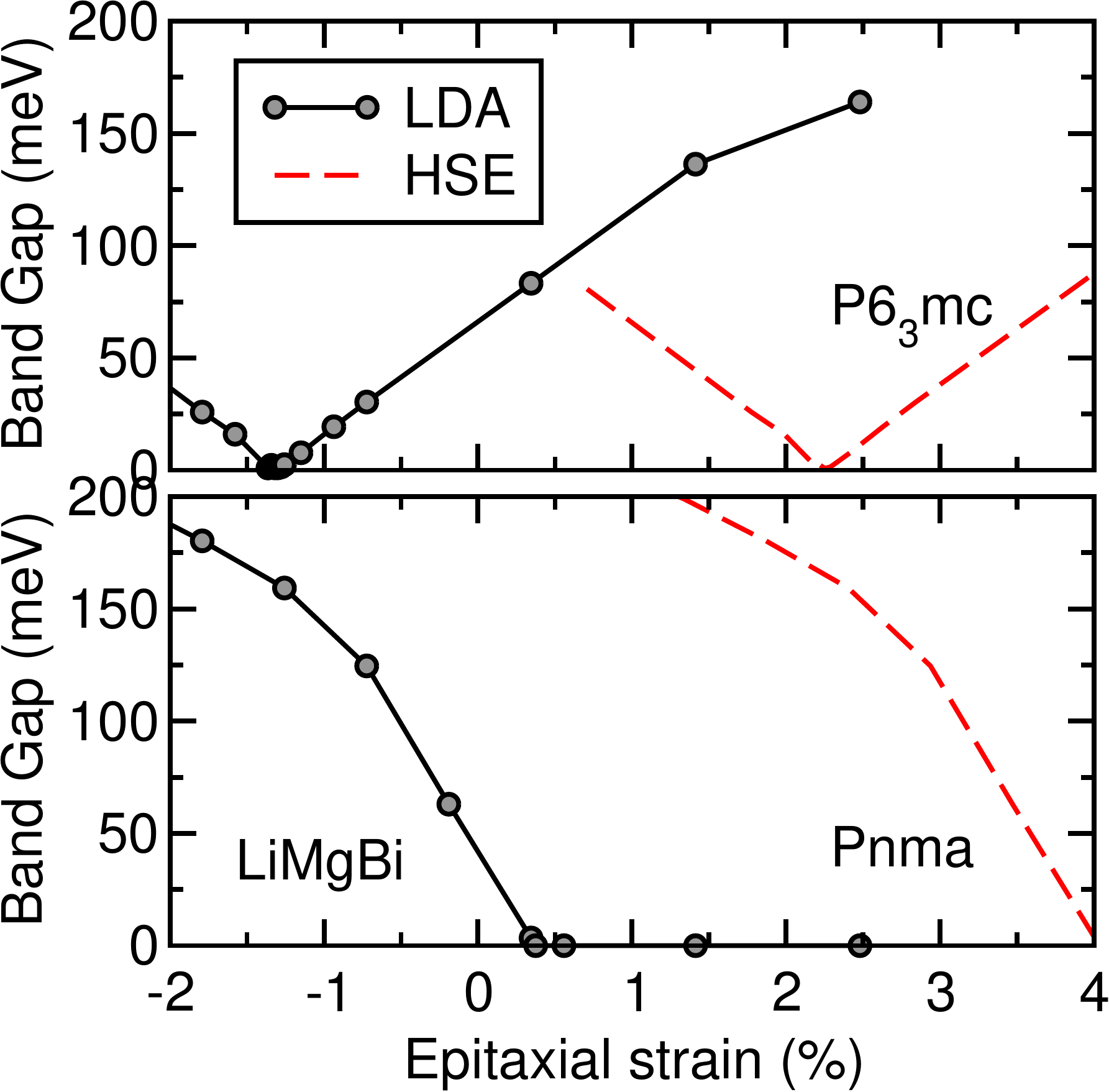}\label{subfig:gap_lmb-hse}}
\hspace{0.6cm}
\subfloat[][]{\includegraphics[scale=0.40]{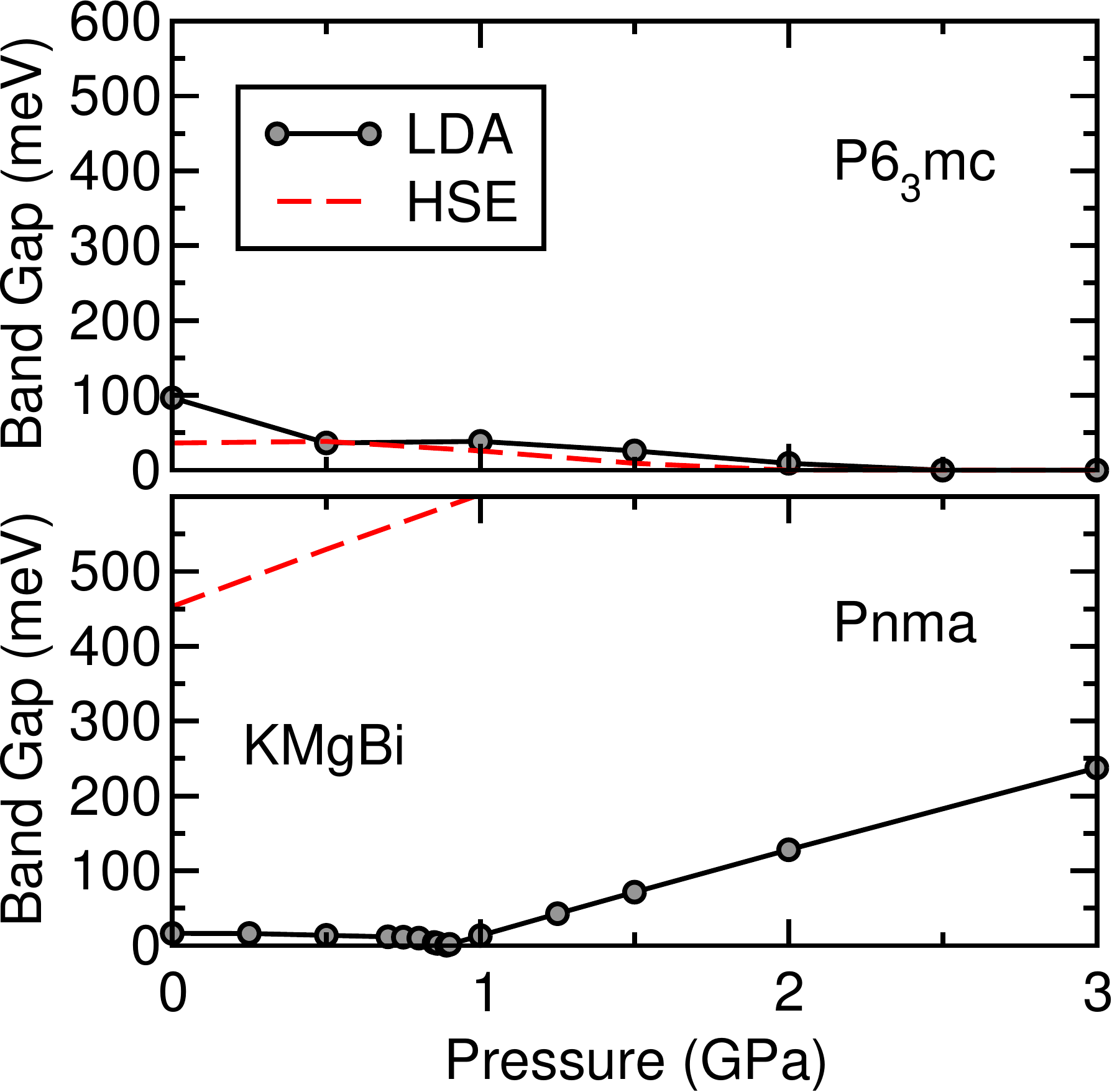}\label{subfig:gap_kmb-hse}}
\caption{Minimum band gap of the LiMgBi (left) and KMgBi (right) compounds of polar $P6_3mc$ (top) and antipolar $Pnma$ (bottom) structures as a function of epitaxial strain and hydrostatic pressure, respectively. The HSE results for the minimum band gap are extrapolated from the band gap at the $\Gamma$ point as described in the text.}
\label{fig:gap-hse}
\end{figure}

To confirm this, we have performed selected band gap calculations using the hybrid HSE functional~\cite{hse06_functional,hse06_functional_erratum}, and the results are shown in Fig.~\ref{fig:gap-hse} for LiMgBi under epitaxial strain (left) and KMgBi under hydrostatic pressure (right) for the polar $P6_3mc$ (top) and antipolar $Pnma$ (bottom) structures. Due to the high computational cost of hybrid functional calculations, we have adopted the following strategy to reduce the number of calculations required, which we exemplify with the $P6_3mc$ structure of LiMgBi. We have performed HSE calculations at strains between $-1.5$\% and $+4.0$\%, but rather than attempting to locate the minimum band gap at each strain, we have instead only calculated the band gap at the $\Gamma$ point. Comparing the minimum and $\Gamma$-point band gaps at the LDA level suggests that the $\Gamma$-point band gap is a good proxy for the behaviour of the minimum band gap. Indeed, the HSE $\Gamma$-point band gap exhibits a minimum at a strain of $+2.3$\%. The $\Gamma$-point band gap at this minimum has a value of $0.12$~eV, and the character of the bands switches between the valence and conduction bands across this minimum. The same behaviour is observed for LDA calculations, but the minimum $\Gamma$-point band gap then occurs at strains around $-1.3$\%, which coincides with the strain-induced topological phase transition. The difference in the $\Gamma$-point band gaps between the LDA and the HSE calculations is constant over the entire strain range, and therefore, for clarity, in Fig.~\ref{fig:gap-hse} we show the HSE results as a dashed line which corresponds to the shifted LDA results by the appropriate amount. A similar procedure is used to obtain the other diagrams shown in Fig.~\ref{fig:gap-hse}. We note that the $P6_3mc$ structure of KMgBi shows a smaller band gap shift than the rest in going from LDA to HSE, and this is caused by a reordering of the valence bands when changing the level of theory used.

%We have then compared the $\Gamma$-point band gaps between the LDA and the HSE calculations, and we expect that these will serve as an appropriate proxy for the trends in the real minimum band gap, as in the LDA calculations the latter is located very close to the $\Gamma$ point. Furthermore, we have confirmed that the HSE band inversions at the $\Gamma$ point are similar to those calculated with the LDA functional. For the HSE calculations we have used the LDA relaxed volumes and internal coordinates.

In LiMgBi (Fig.~\ref{subfig:gap_lmb-hse}), the HSE band gap inversion for the polar $P6_3mc$ structure occurs at an epitaxial strain of about $+2.3\%$, compared to $-1.3\%$ for LDA, which corresponds to a shift in epitaxial strain of about $3.6$\%. Similarly, the epitaxial strain at which the antipolar $Pnma$ structure becomes metallic increases from the LDA $+0.4\%$ value to about $+4.0\%$ for HSE, a similar increase of $3.6$\% in epitaxial strain. These results imply that LiMgBi exhibits an AFTI phase of type I at strains in the range $+2.3$ to $+4.0$\% in the HSE description.

In KMgBi (Fig.~\ref{subfig:gap_kmb-hse}), we predict that for the ground state antipolar $Pnma$ structure, the material is in the normal phase already at $0$~GPa, and the topological phase is pushed to \textit{negative} pressures. For the polar $P6_3mc$ structure, the transition to the metallic phase occurs at pressures about $0.5$~GPa lower using the HSE functional compared to the LDA functional, and the material stays in the topological phase for pressures below about $2$~GPa. These results imply that KMgBi is an AFTI of type I under ambient conditions, and remains so at moderate pressures.

Overall, the use of a hybrid functional leads to quantitative differences in the phase diagrams of LiMgBi under strain and KMgBi under pressure. But the general structure of the phase diagrams remains unchanged. In particular, the existence of an AFTI of type I in both materials is robust with respect to the choice of exchange-correlation functional.

\end{document}